\documentclass[conference]{IEEEtran}
\IEEEoverridecommandlockouts
\usepackage{cite}
\usepackage{amsmath,amssymb,amsfonts}
\usepackage{algorithmic}
\usepackage{graphicx}
\usepackage{textcomp}
\usepackage{xcolor}
\usepackage{pgf}
\usepackage{cleveref}
\usepackage{booktabs}
\usepackage{stfloats}       
\usepackage[section]{placeins} 
\usepackage[most]{tcolorbox}
\usepackage{subcaption}       

\newtcolorbox{takeaway}{%
    enhanced,
    colback=black!4,
    colframe=black!45,
    boxrule=0.4pt,
    arc=1.5pt,
    left=6pt, right=6pt, top=5pt, bottom=5pt,
    fontupper=\small,
    before upper={\textit{\textbf{Takeaway.}}\enspace},
}
\def\BibTeX{{\rm B\kern-.05em{\sc i\kern-.025em b}\kern-.08em
    T\kern-.1667em\lower.7ex\hbox{E}\kern-.125emX}}
\begin{document}

\title{Design Methodology and Performance Trade-offs Management for Distributed and Compound AI Systems}

\author{
  \IEEEauthorblockN{Milos Gravara}
  \IEEEauthorblockA{
    Distributed Systems Group\\
    TU Wien\\
    m.gravara@dsg.tuwien.ac.at
  }\and
  \IEEEauthorblockN{Andrija Stanisic}
  \IEEEauthorblockA{
    Distributed Systems Group\\
    TU Wien\\
    a.stanisic@dsg.tuwien.ac.at
  }
  \and
  \IEEEauthorblockN{Stefan Nastic}
  \IEEEauthorblockA{
    Distributed Systems Group\\
    TU Wien\\
    s.nastic@dsg.tuwien.ac.at
  }
}

\maketitle

\begin{abstract}
    Artificial Intelligence (AI) systems must typically satisfy service-level objectives including accuracy, latency, and cost. The prevailing model-centric approaches select a monolithic model at design time and use it to process all inputs. Monolithic models apply identical computation regardless of input difficulty, cannot decompose tasks across specialized components, and have knowledge that is fixed at training time. During runtime, this can lead to performance degradation and increasing costs. Because the model is the most important design variable, it determines the majority of system behavior, coupling operational objectives to a single design-time choice. Addressing these limitations requires shifting from model-centric to system-centric design. Compound AI systems realize this shift by orchestrating multiple models, algorithms, and tools as distributed AI systems through explicit control logic. The performance of such systems depends on their workflow topology, the models assigned to each task, and the parameters governing runtime behavior. We present a design methodology that organizes this space along two dimensions, workflow topology and configuration selection, and identifies eight design patterns, each consolidating techniques to address a specific limitation of monolithic deployment. We validate our methodology through three case studies spanning object detection, question answering, and mathematical reasoning. Across our case studies, Compound AI configurations approach accuracy of monolithic models within 2.5 to 4 percentage points while reducing latency by up to 60\% and cost by up to 71\%. We show that model selection and parameter configuration jointly determine system performance, but the resulting design space grows combinatorially, as workflows compose more patterns and components. Thus, we identify five open challenges that define a roadmap from manually configured prototypes towards systems that automatically discover and maintain SLO-compliance in Compound and Distributed AI systems.
\end{abstract}


\section{Introduction}\label{sec:intro}

Artificial intelligence (AI) systems are deployed at massive scale across applications such as video analytics, autonomous navigation, conversational assistants, and code generation~\cite{bommasani2022opportunities}. Advances in model architectures and training methods~\cite{vaswani2017attention, brown2020language, hoffmann2022training} have expanded their capabilities to include complex reasoning, multimodal understanding, and open-end interactions. As adoption grows, these models must satisfy strict service level objectives (SLOs), which typically include latency, throughput, and cost. This places increasing pressure on how models are deployed~\cite{miao2025efficient, golec2023reconciling}.

The prevailing approach for AI deployment is model-centric~\cite{zaharia2024shift, miao2025efficient}. A single, monolithic model is selected at design time based on task-specific benchmarks and effectively remains fixed during serving. Every input follows the same computational path and consumes the same resources regardless of its content or difficulty. Improving performance for a given task relies primarily on scaling the model through additional parameters, training data, or compute budget, following empirical scaling laws~\cite{kaplan2020scaling, hoffmann2022training, LLMsTooBig}.

However, model-centric AI systems exhibit structural limitations that cannot be resolved through scaling alone~\cite{srivastava}. First, monolithic models apply the same inference process to every request regardless of its difficulty and domain. As a result, simple queries consume the same computational resources as complex ones, and there is no mechanism to decompose tasks across specialized models~\cite{chen2023frugalgpt, ding2024hybridllm}. Second, many real-world tasks require capabilities beyond model inference, such as precise computation, interaction with external systems, or output verification. These operations cannot be reliably performed by single models, regardless of their size~\cite{schick2023toolformer}. Third, the knowledge encoded in model parameters is fixed at training time. Incrementally updating this knowledge through retraining or fine-tuning becomes prohibitively expensive as model scale increases~\cite{RAISINGCOSTS}.

In production, these structural limitations directly affect SLOs~\cite{kapoor2024agents}. Because the model is the most important design variable, it determines the majority of system behavior, coupling operational objectives to choices made at design time~\cite{chen2025modelselection}, with no mechanism to control them independently. Satisfying SLOs requires selectively allocating computation across requests, composing specialized components for different aspects of a task, and adapting system behavior to changing workloads. A monolithic model whose behavior is fixed at inference time cannot provide this control. Addressing these requirements demands a shift from model-centric to system-centric approach, where the unit of design is no longer the model but the system that orchestrates models, algorithms, and tools.

Compound AI~\cite{zaharia2024shift} is an emerging paradigm which attempts to address these limitations by shifting the design focus from a single model to a composed system. Compound AI systems combine specialized AI models and engineered software components into workflows with explicit control logic. Instead of relying on a single model, the system can retrieve external knowledge~\cite{lewis2020retrieval}, route requests to models with different capabilities~\cite{chen2023frugalgpt, ding2024hybridllm}, invoke tools for specialized tasks~\cite{schick2023toolformer}, and allocate additional inference-time computation when necessary~\cite{wang2023selfconsistency}. However, the performance of such systems no longer depends on a single model choice, but on multiple design decisions, such as the topology of the workflow, the assignment of models to each inference task, and the parameters that govern runtime behavior~\cite{COMPASS, OPTIMAS}.

In this paper, we present a design methodology for Compound AI systems that decomposes system design into two dimensions: workflow structure and configuration selection. We identify eight design patterns that capture common workflow topologies addressing key limitations of monolithic deployment. We evaluate the methodology across three case studies in vision and language domains, demonstrating that Compound AI systems can approach monolithic baselines while improving latency and cost. 

Our main contributions include:  

\begin{itemize}
    \item We consolidate recurring workflow composition techniques into eight design patterns for Compound AI systems, each addressing a specific limitation of monolithic deployment.
    \item We empirically characterize Compound AI performance across three Compound AI workflows in vision and language domains. Throughout experiments, Compound AI systems achieve task performance comparable to monolithic baselines, while yielding efficiency gains of up to 60\% in latency and 71\% in cost. 
    \item We identify five open challenges in Compound AI that define a roadmap from manually configured prototypes towards AI systems that automatically discover and maintain SLO-compliance.
\end{itemize}

The rest of the paper is organized as follows. Section~\ref{sec:motivation} provides background on monolithic AI deployment and its limitations. Section~\ref{sec:refarch} presents the design methodology, including design patterns and configuration selection. Section~\ref{sec:trade} validates the methodology through case studies and identifies key Compound AI properties. Based on these findings, Section~\ref{sec:challenges} lays out future research challenges. Section~\ref{sec:related} discusses related work, and Section~\ref{sec:conclusion} concludes the work.

\section{Motivation}\label{sec:motivation}

This section motivates the shift from model-centric to system-centric AI approaches. Section~\ref{sec:motivation:monolithic} establishes that in monolithic deployment model choice is the most important design variable, with model scaling as the only mechanism for improvement~\cite{kaplan2020scaling, hoffmann2022training}. Section~\ref{sec:motivation:serving} shows that production deployment fixes all three SLO dimensions to a single operating point that inference serving cannot change. Section~\ref{sec:motivation:limitations} demonstrates that satisfying production SLOs requires degrees of freedom that a single model cannot provide.

\subsection{Monolithic AI Models} \label{sec:motivation:monolithic}

The standard approach to AI deployment serves a single model as the
complete inference pipeline. A monolithic deployment selects one model with fixed parameters and applies it uniformly to all incoming requests. The model is chosen at design time based on task-specific benchmarks, and its parameters remain fixed throughout serving. Every input, regardless of its content or difficulty, follows the same computational path and consumes the same resources. The choice of model is therefore the most important design variable in this paradigm, simultaneously determining output quality, computational cost, and serving latency.

The primary mechanism for improving monolithic model performance is
scaling, which involves increasing the number of parameters, the volume of training data, or the training compute budget. Scaling laws establish that pretraining loss decreases as a power law with respect to model size, dataset size, and compute~\cite{kaplan2020scaling, hoffmann2022training}. This is a \textit{model-centric} paradigm that has driven consistent progress across tasks and modalities, with improvements measured by accuracy, F1, or similar metrics on held-out benchmarks. 

\subsection{Inference Serving} \label{sec:motivation:serving}

A model in production must satisfy SLOs spanning accuracy, latency, and cost~\cite{kapoor2024agents}. These objectives are coupled: higher accuracy consistently requires greater computational cost and latency. Inference serving systems manage deployments through request batching, memory allocation, and workload scheduling, with modern frameworks establishing standard practices through continuous batching, fragmentation-aware memory management, and hardware-aware scheduling~\cite{kwon2023vllm, yu2022orca, miao2024towards}. Such optimizations make a fixed model run more efficiently, but cannot change which model processes a given input or how much computation it receives. Because the model is the sole component of the deployed system, the model choice determines where the system operates across all three SLO dimensions. One model produces one operating point in the accuracy-latency-cost space, fixed at design time and unchanged throughout deployment.

\subsection{Limitations of Monolithic Deployment} \label{sec:motivation:limitations}

\begin{figure}[t]
    \centering
    \includegraphics[width=\columnwidth]{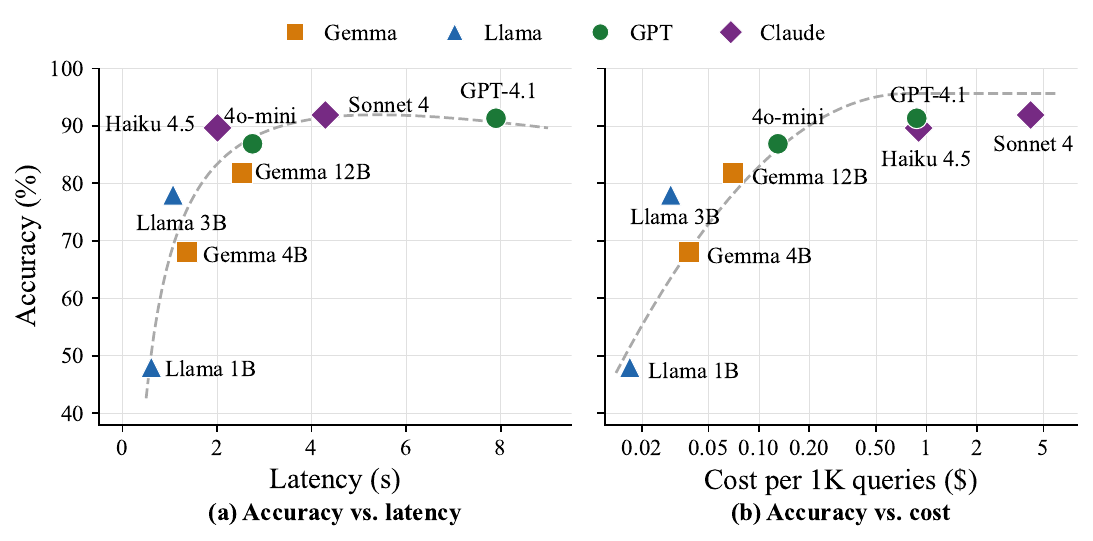}
    \caption{Monolithic operating points on GSM8K (100 problems). Each model occupies a fixed position in the (a)~accuracy-latency and (b)~accuracy-cost planes. }
    \label{fig:monolithic}
\end{figure}
\Cref{fig:monolithic} illustrates this coupling on GSM8K~\cite{cobbe2021verifiers} benchmark across eight models from four model families: Gemma and Llama, served locally on an A4000 GPU (16GB), and GPT and Claude, accessed through proprietary APIs. Each model occupies a single point in both the accuracy-latency plane (\Cref{fig:monolithic}a) and the accuracy-cost plane (\Cref{fig:monolithic}b). The trend in both panels shows that moving toward higher accuracy requires accepting higher latency and higher cost simultaneously. No model reaches the top-left region of either panel, where high accuracy would coincide with low latency or low cost. The design variable is the model itself, and each model produces exactly one operating point with no mechanism to trade off one objective against another at runtime.

This rigidity is particularly wasteful because workloads are
non-uniform. Inputs vary in difficulty, domain, and computational
requirements. On GSM8K, for example, cheap local models such as
Llama3.2-3B already solve 78\% of problems correctly, while the remaining fraction requires models that cost up to two orders of magnitude more per query. Monolithic deployment ignores this structure by applying a single model uniformly to all inputs, over-provisioning for easy inputs and under-provisioning for hard ones. 


Beyond efficiency, monolithic models have structural limitations that scaling cannot resolve. A model's knowledge is determined entirely by its training data, so it cannot access information that emerged after training without retraining or fine-tuning. This can turn out to be prohibitive, as the cost of incremental improvement grows exponentially~\cite{RAISINGCOSTS}. Additionally, a monolithic model must cover every capability the task demands, even when dedicated models trained for specific aspects of the task would produce better results~\cite{jiang2023llmblender}. These are architectural rather than capacity constraints, as adding parameters does not give a model the ability to access external knowledge, or delegate work to a more suitable model.

Meeting these requirements demands internal degrees of freedom that monolithic deployments cannot provide. Specifically, a system needs the ability to decide which model handles which input, when a result is sufficient to stop further computation, how much external knowledge to incorporate for a given request, whether an output meets quality criteria before it is served, and how to apportion computation across components when resources are constrained. None of these decisions can be expressed within a monolithic deployment because there is no point at which the system can observe input properties, intermediate results, or operating conditions and act on them. This is why addressing these limitations requires moving from model-centric to system-centric design, where the composition of models, algorithms, and tools becomes the design object and the decisions above become explicit, configurable aspects of the system.

\section{Design Methodology for Compound AI}\label{sec:refarch}









We propose a design methodology for Compound AI systems. Compound AI systems are distributed AI systems that orchestrate AI models and software-engineered components through explicit control logic to meet service-level objectives. The control logic determines which component processes a given input, what context it receives, and how intermediate results are combined.

Compound AI systems can be conceptually described using three main types of building blocks. \textit{AI Models} perform the core inference tasks such as prediction, generation, or classification. \textit{Algorithms} provide deterministic logic for control and coordination: dispatching requests, evaluating output quality, combining results, and enforcing constraints. \textit{Tools} provide external capabilities such as databases, search indices, or code execution that extend the system beyond what models and algorithms can compute. These categories determine what types of parameters each component exposes, as discussed in Section~\ref{subsec:config}.

The design methodology decomposes Compound AI system design into two dimensions. The first dimension is \textbf{workflow topology}, which determines how components are arranged and how data flows between them. \Cref{subsec:patterns} identifies eight design patterns that each address a specific limitation of monolithic deployments. The second dimension is \textbf{configuration selection}, which assigns specific models to inference tasks and sets the parameters that govern runtime behavior. \Cref{subsec:config} shows that this choice determines where the system operates within the accuracy-latency-cost space. 

\begin{figure*}[!tp]
\centering

\includegraphics[width=0.5\linewidth]{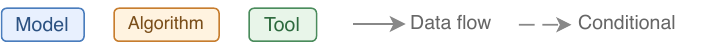}

\vspace{2pt}
\hrule height 0.2pt
\vspace{4pt}

\begin{subfigure}[t]{0.235\textwidth}
    \centering
    \includegraphics[width=\linewidth]{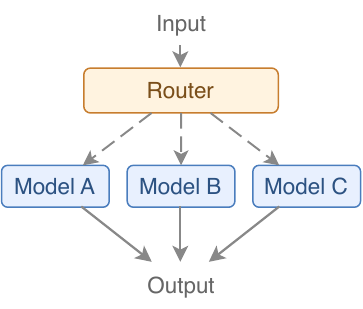}
    \caption{Router}
    \label{fig:pat:router}
\end{subfigure}%
\hfill
\begin{subfigure}[t]{0.235\textwidth}
    \centering
    \includegraphics[width=\linewidth]{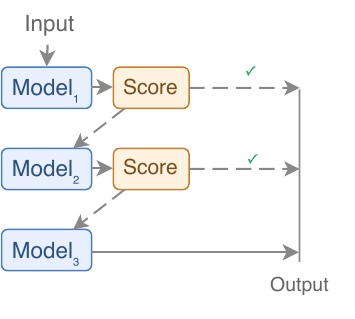}
    \caption{Cascade}
    \label{fig:pat:cascade}
\end{subfigure}%
\hfill
\begin{subfigure}[t]{0.235\textwidth}
    \centering
    \includegraphics[width=\linewidth]{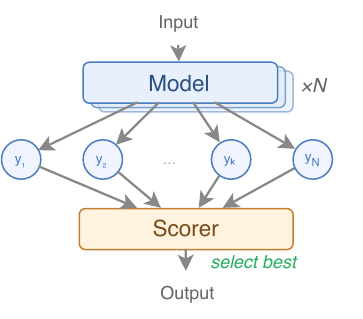}
    \caption{Sampler}
    \label{fig:pat:sampler}
\end{subfigure}%
\hfill
\begin{subfigure}[t]{0.235\textwidth}
    \centering
    \includegraphics[width=\linewidth]{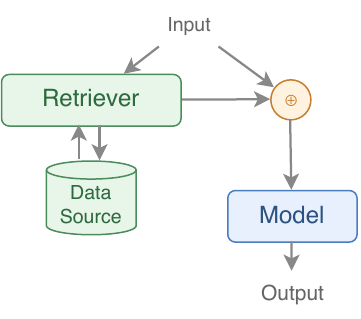}
    \caption{Retriever}
    \label{fig:pat:retriever}
\end{subfigure}

\vspace{6pt}

\begin{subfigure}[t]{0.235\textwidth}
    \centering
    \includegraphics[width=\linewidth]{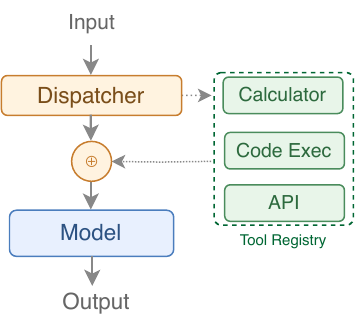}
    \caption{Tool Executor}
    \label{fig:pat:toolexecutor}
\end{subfigure}%
\hfill
\begin{subfigure}[t]{0.235\textwidth}
    \centering
    \includegraphics[width=\linewidth]{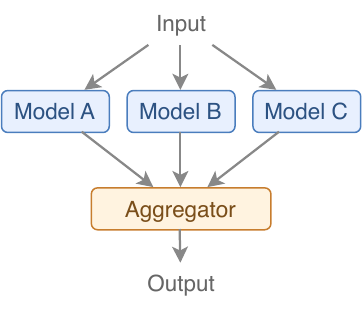}
    \caption{Aggregator}
    \label{fig:pat:aggregator}
\end{subfigure}%
\hfill
\begin{subfigure}[t]{0.235\textwidth}
    \centering
    \includegraphics[width=\linewidth]{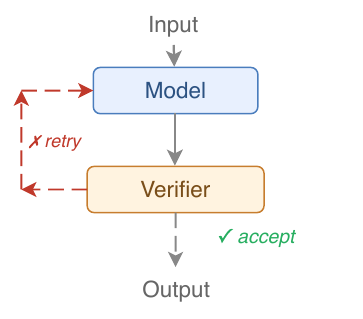}
    \caption{Verifier}
    \label{fig:pat:verifier}
\end{subfigure}%
\hfill
\begin{subfigure}[t]{0.235\textwidth}
    \centering
    \includegraphics[width=\linewidth]{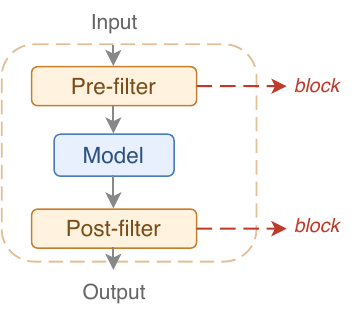}
    \caption{Guardrail}
    \label{fig:pat:guardrail}
\end{subfigure}

\vspace{2pt}
\caption{Design pattern topologies.}
\label{fig:patterns}

\vspace{4pt}
\hrule height 0.4pt
\vspace{4pt}

\setlength{\tabcolsep}{3pt}
\scriptsize
\renewcommand{\arraystretch}{1.15}
\begin{tabular*}{\linewidth}{@{}p{0.09\linewidth} p{0.21\linewidth} p{0.22\linewidth} p{0.18\linewidth} p{0.20\linewidth}@{}}
\textbf{Pattern} & \textbf{Problem} & \textbf{Solution} & \textbf{Expected Effect} & \textbf{Example} \\
Router &
Monolithic deployment applies the same model to every request, wasting cost and latency on inputs that a cheaper model could resolve correctly. &
A control component (e.g., algorithm or learned model) evaluates input properties and dispatches each request to an appropriately sized model. &
Reduces cost and latency; risks accuracy loss when the router misclassifies input difficulty &
RouteLLM~\cite{ong2024routellm}, Hybrid LLM~\cite{ding2024hybridllm}, ECO-LLM~\cite{patidar2025ecollm}, collaborative object detection for edge video~\cite{khani2021realtime, COLLABORATIVEOD, VATE}. \\
\addlinespace
Cascade &
Monolithic deployment applies the same model to every request, but input difficulty cannot be reliably estimated before inference. &
Models are invoked sequentially from lightweight to heavyweight, with a scoring function (algorithm or model) evaluating output quality at each step; processing stops when quality is sufficient. &
Reduces average cost by resolving most requests cheaply; increases worst-case latency when escalation is required. &
FrugalGPT~\cite{chen2023frugalgpt}, speculative cascades~\cite{narasimhan2024speculative}, AutoMix~\cite{aggarwal2024automix}, Viola-Jones attentional cascade~\cite{viola2001rapid}, MTCNN~\cite{zhang2016joint}, BranchyNet early exits~\cite{teerapittayanon2016branchynet}. \\
\addlinespace
Sampler &
A monolithic model's output quality is bounded by a single inference pass, with no mechanism to improve quality by investing additional compute at inference time. &
One or more models generate $N$ candidate outputs via repeated sampling; an aggregation component combines or selects among candidates to produce final output. &
Improves output quality by scaling inference-time compute; cost increases linearly with $N$. &
Self-consistency~\cite{wang2023selfconsistency}, best-of-$N$~\cite{davis2024networks}, AlphaCode~\cite{li2022competition}, scaling test-time compute~\cite{snell2024scaling, balachandran2025scaling}, multi-agent debate~\cite{du2024improving}, vote and filter-vote~\cite{chen2024scaling}. \\
\addlinespace
Retriever &
A monolithic model's knowledge is fixed at training time, and updating it through retraining becomes prohibitively expensive as model scale increases. &
A retrieval tool (e.g., database or search index) provides relevant external context that is merged into the model's input before generation. &
Improves accuracy and knowledge coverage without scaling model size; adds retrieval latency and depends on retrieval quality. &
RAG~\cite{lewis2020retrieval}, REALM~\cite{guu2020realm}, VideoRAG~\cite{ren2025videorag}, MedRAG~\cite{zhao2025medrag}. \\
\addlinespace
Tool \newline Executor &
A monolithic model cannot reliably perform precise computation, execute code, or interact with external systems. &
Subtasks requiring deterministic execution are delegated to external tools such as calculators, code interpreters, or API endpoints. &
Improves correctness on tasks requiring precise computation or external interaction. Introduces dependency on tool availability. &
Toolformer~\cite{schick2023toolformer},  function calling~\cite{gpt}, code interpreters~\cite{gpt}, MCP tool servers~\cite{MCP}. \\
\addlinespace
Aggregator &
A single model produces one output per request with no mechanism to pool predictions across multiple models or input partitions running in parallel, leaving accuracy bounded by what one model can produce alone. &
Multiple models or instances process the same input or different partitions of it in parallel; an aggregation component combines outputs through, for example, voting, fusion, or merging. &
Improvement in accuracy and robustness through complementary predictions; support for parallel processing of large inputs; increased cost with the number of models or partitions. &
Vote and Filter-Vote~\cite{chen2024scaling}, multi-agent debate~\cite{du2024improving}, ChatEval~\cite{chan2023chateval}, Weighted Boxes Fusion for detection~\cite{solovyev2021weighted}, Random Forests~\cite{breiman2001random}. \\
\addlinespace
Verifier &
Single-pass generation provides no output quality guarantees. Monolithic model cannot detect or correct its own errors before the output is served. &
A verification component (e.g., algorithm or model) evaluates output against acceptance criteria; the system re-generates outputs that fail with modified parameters up to a retry budget. &
Improvement in output reliability through iterative refinement; increased latency and cost scaling with the number of retries. &
Training verifiers for math~\cite{cobbe2021verifiers}, LEVER execution-based code verification~\cite{ni2023lever}, Self-Refine~\cite{madaan2023selfrefine}, AlphaCodium test-based verification~\cite{ridnik2024alphacodium}, generator-verifier networks~\cite{davis2024networks}. \\
\addlinespace
Guardrail &
A monolithic model has no built-in mechanism to enforce safety or compliance constraints independently of its own output, making safety guarantees dependent on model behavior. &
Pre- and post-processing filters (e.g., algorithms or specialized models) enforce safety and compliance constraints around the core model. &
Enforcement of safety independent of model behavior; increased latency and potential utility degradation due to over-filtering. &
NeMo Guardrails~\cite{rebedea2023nemo}, Llama Guard~\cite{inan2023llamaguard}, TorchOpera~\cite{han2024torchopera}, SafeWatch video guardrails~\cite{chen2025safewatch}. \\
\hline
\end{tabular*}

\captionof{table}{Design patterns for Compound AI systems. Each pattern addresses a limitation of monolithic deployment through a specific structural solution, producing a characteristic effect on system objectives.}
\label{tab:patterns}

\end{figure*}

\subsection{Design Patterns}\label{subsec:patterns}

Each design pattern defines a workflow topology that addresses a specific limitation of monolithic deployment. These patterns generalize techniques already established in practice, referenced in the Example column of Table~\ref{tab:patterns}, recast as reusable structural abstractions. Fig.~\ref{fig:patterns} illustrates the topology of each pattern, and Table~\ref{tab:patterns} summarizes the problem that each pattern addresses, its structural solution, the expected effect on system objectives, and examples. We organize the eight patterns into four groups by the objective they primarily target: compute efficiency, output quality, capability extension, and safety.

\textit{Compute Efficiency.} The Router and Cascade patterns can improve compute efficiency by avoiding invocation of the same model for every request, regardless of its difficulty. Both patterns selectively allocate inference across models of different capability, but differ in 
timing of the allocation decision. The Router evaluates input properties, such as input difficulty or domain, before inference and dispatches each request to an appropriate model. The Router commits to a single model upfront, which is cheaper when the estimate is correct but risks misrouting. The Cascade, however, invokes models sequentially from lightweight to heavyweight and stops when a  scoring function determines the output is sufficient. The Cascade always starts cheap and escalates only when necessary, but may invoke multiple models for hard inputs.

\textit{Output quality.} The Sampler, Verifier, and Aggregator patterns can be used to improve output quality beyond what a single inference pass achieves. The Sampler generates $N$ candidate outputs and selects the best through voting or scoring, trading additional compute for quality. The Verifier evaluates inference output against acceptance criteria and re-triggers inference when quality is insufficient, refining the output through iteration. The Aggregator pattern runs multiple models or model instances in parallel and combines their outputs through voting, fusion or merging. All three patterns provide a mechanism to invest additional compute to gain output quality. This opens the possibility of using smaller, more efficient models that recover quality through repeated or combined inference rather than model scale.

\textit{Capability Extension.} The Retriever and Tool Executor can extend the system with capabilities that model inference alone cannot provide. The Retriever may augment the model's input with relevant external context fetched from a database or search index, addressing the limitation that model knowledge is fixed at training time. The Tool Executor can delegate subtasks requiring deterministic execution to external tools such as code interpreters or API endpoints, addressing the limitation that models cannot reliably perform precise computation. Both patterns allow the system to evolve its capabilities independently of the model, by updating retrieval sources or adding new tools without retraining.

\textit{Safety and compliance.} The Guardrail pattern can enable safety as a system-level property rather than a model-level behavior. Pre-processing and post-processing filters enforce safety and compliance constraints independently of the model that performs the task. This allows safety guarantees to hold across different model configurations and to be audited and updated without modifying the underlying model.

Each pattern addresses a specific limitation of monolithic deployment, but introduces trade-offs in doing so. A Router can significantly reduce cost and latency, but risks accuracy loss when the routing decision is incorrect. A Retriever may improve accuracy through external context, but also may introduce latency in the process of retrieving relevant information. In practice, patterns compose to balance these trade-offs. A system that routes inputs to cheaper models can recover accuracy through retrieval, combining both patterns into a single workflow topology. The resulting topology specifies structural roles and control flow, but is independent of which specific models, algorithms, or tools fill each role. Assigning concrete components to these roles and setting the parameters that govern their runtime behavior is the subject of configuration selection.

\subsection{Configuration Selection}
\label{subsec:config}

The topology alone does not determine the Compound AI system behavior. For instance, two systems with identical topologies can produce different accuracy, latency, and cost if they use different models or different parameter settings. Configuration selection assigns specific models to the inference roles defined by the topology and sets the parameters that govern runtime behavior. We define a \textit{system configuration} as a complete assignment of models to all inference roles together with values for all parameters. A configuration fully specifies how the system processes any given input. Changing any element of the configuration, whether a model assignment or a single parameter value, can shift the system's operating point across all performance dimensions.

Model selection assigns a model to each inference role in the workflow topology. A Router pattern, for instance, requires a routing mechanism and a set of candidate models for dispatch. The choice of models determines the accuracy each path can achieve, the latency each path incurs, and the cost of each inference call, making model selection the most consequential configuration dimension. Similar decisions apply to algorithms and tools, such as which routing mechanism or which retrieval index to use, though these choices are typically more constrained by the pattern structure and deployment environment.


Parameter configuration sets the values that govern runtime behavior within a fixed topology and model assignment. These include routing thresholds that determine when a request is dispatched to a cheaper model, retrieval depth that controls how much external context is provided, sample counts that govern how many candidate outputs are generated, and model-level settings such as temperature or batch size. Different parameter values shift the system's operating point across accuracy, latency, and cost.

\section{Case Studies in Compound AI systems}\label{sec:trade}

We validate our design methodology through three case studies that instantiate design patterns from Section~\ref{sec:refarch} across vision and language domains. Each study compares Compound AI performance against monolithic baselines and characterizes trade-offs between task performance, latency, and cost.


\subsection{Experimental Setup}\label{subsec:setup}

\begin{figure*}[t]
\centering
\begin{subfigure}[t]{0.32\textwidth}
    \centering
    \includegraphics[width=\linewidth]{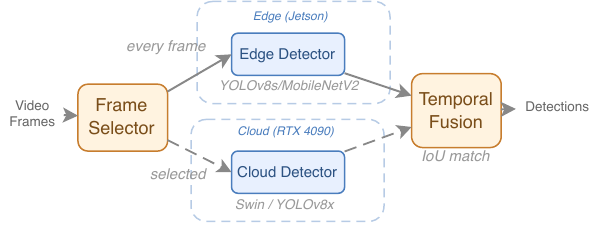}
    \caption{CODEC}
    \label{fig:wf:codec}
\end{subfigure}%
\hfill
\begin{subfigure}[t]{0.32\textwidth}
    \centering
    \includegraphics[width=\linewidth]{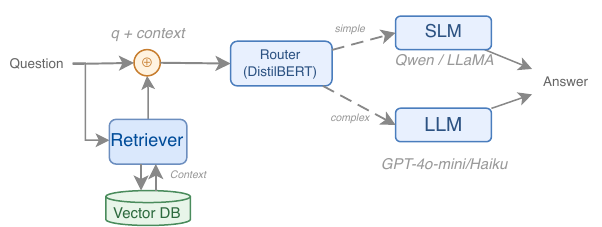}
    \caption{QARouter}
    \label{fig:wf:ragrouter}
\end{subfigure}%
\hfill
\begin{subfigure}[t]{0.32\textwidth}
    \centering
    \includegraphics[width=\linewidth]{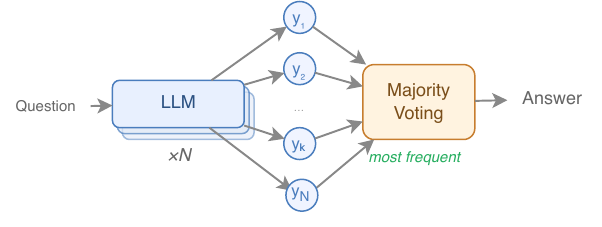}
    \caption{InferScale}
    \label{fig:wf:scalellm}
\end{subfigure}

\caption{Workflow topologies of analyzed use cases.}
\label{fig:workflows}
\end{figure*}

Fig.~\ref{fig:workflows} shows the three workflow topologies. CODEC is a collaborative edge-cloud object detection system that processes streaming drone video. A lightweight detection model runs on every incoming frame at the edge, while a more capable detection model runs in the cloud on selected frames sent asynchronously. A temporal fusion algorithm merges cloud predictions with edge detections using IoU matching. CODEC instantiates the Router and Aggregator patterns.

QARouter combines retrieval-augmented generation with learned routing for question answering. Every incoming query first retrieves relevant passages from a vector database. A trained classifier then evaluates the query alongside the retrieved context and dispatches it to either a local or a proprietary LLM for generation. QARouter composes the Retriever and Router patterns. 

InferScale applies test-time compute scaling for reasoning tasks. For each input, the system generates $N$ candidate answers using repeated sampling and selects the best response through majority voting. Thus, InferScale instantiates the Sampler pattern, with the sample count $N$ as the primary configuration parameter.

The case studies are implemented in Python 3.10 as research prototypes\footnote{https://github.com/polaris-slo-cloud/compound-ai}. Experiments run on a testbed with an NVIDIA Jetson Orin Nano (8GB) and NVIDIA A4000 (16GB) for lightweight inference, and an NVIDIA RTX 4090 (24GB) for large-model serving and local LLM inference via Ollama. Each workflow component executes in a Docker container with NVIDIA runtime, ensuring reproducible resource allocation. Proprietary models are accessed through public APIs.

\begin{table}[t]
\centering
\caption{Experimental configuration for case studies.}
\label{tab:setup}
\scriptsize
\renewcommand{\arraystretch}{1.3}
\begin{tabular}{@{}l p{0.8\columnwidth}@{}}
\toprule
\multicolumn{2}{@{}l}{\textbf{CODEC}} \\
\midrule
Models         & YOLOv8~\cite{jocher2023yolov8}, MobileNetV2-SSD~\cite{sandler2018mobilenetv2,liu2019ssd}, Swin~\cite{liu2021swin} \\
Dataset        & VisDrone2019-VID~\cite{zhu2021visdrone} \\
Metric         & mAP@50 \\
Baseline       & Swin, YOLOv8x \\
Infrastructure & Jetson Orin Nano 8GB (edge), RTX 4090 24GB (cloud) \\
\addlinespace
\multicolumn{2}{@{}l}{\textbf{QARouter}} \\
\midrule
Models         & DistilBERT~\cite{sanh2020distilbertdistilledversionbert} (router), LLaMA3.2:3B~\cite{Llama3}, Qwen2.5:1.5B~\cite{bai2023qwentechnicalreport}, \\
               & GPT-4o-mini~\cite{gpt}, Claude-3.5-Haiku~\cite{claude} \\
Dataset        & SQuAD v2.0~\cite{SQUADV2} \\
Metric         & F1 \\
Baseline       & GPT-4o-mini, Claude-3.5-Haiku \\
Infrastructure & RTX A4000 16GB, public API \\
\addlinespace
\multicolumn{2}{@{}l}{\textbf{InferScale}} \\
\midrule
Models         & LLaMA3.2:3B (local) \\
Dataset        & GSM8K~\cite{cobbe2021verifiers} \\
Metric         & Accuracy \\
Baseline       & GPT-4o-mini \\
Infrastructure & RTX 4090 24GB, public API \\
\bottomrule
\end{tabular}
\end{table}

We evaluate three dimensions. \textit{Accuracy} captures domain-specific metrics (mAP@50 for detection; F1 for QA; accuracy for mathematical reasoning). \textit{Latency} measures average end-to-end response time. \textit{Cost} estimates per-1K-request expense, combining amortized local GPU cost (Jetson Orin Nano and RTX A4000: \$0.10/hr and RTX 4090: \$0.40/hr~\cite{lambdalabs}) with API token pricing derived from proprietary server vendors~\cite{openai_pricing, anthropic_pricing}. Table~\ref{tab:setup} summarizes the experimental framework, including models, datasets, and baselines.


The evaluation has two objectives: 1) to examine whether Compound AI configurations can approach monolithic accuracy while reducing latency and cost, and 2) to characterize how model selection and parameter configuration shape the accuracy-latency-cost trade-off.


\subsection{CODEC: Edge-Cloud Object Detection}\label{subsec:codec}

\begin{figure*}
    \centering
    \includegraphics[width=0.9\textwidth]{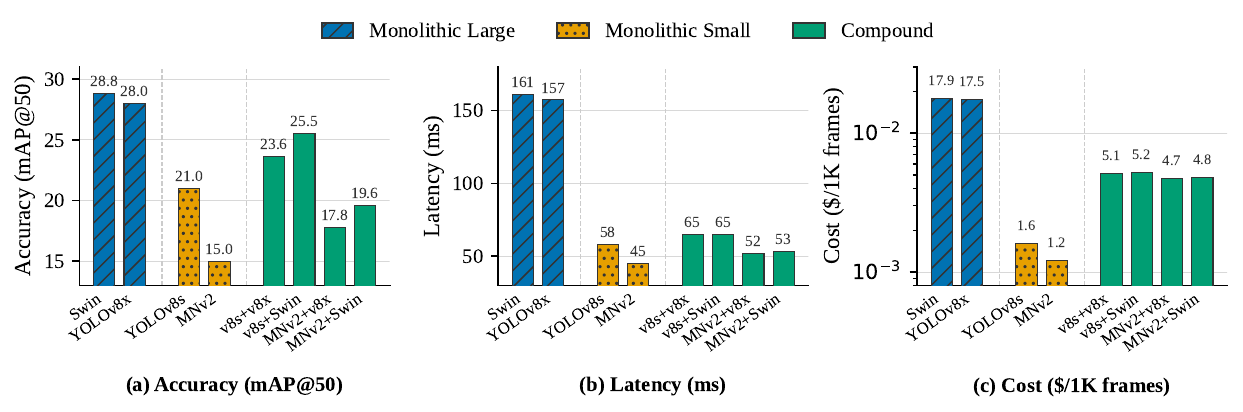}
    \caption{CODEC performance (accuracy, latency, and cost) compared to monolithic baselines on VisDrone dataset.}
    \label{fig:codec_results}
\end{figure*}

CODEC instantiates the Router and Aggregator patterns for streaming drone video. The Router dispatches frames between edge and cloud models based on offloading frequency, while the Aggregator merges their predictions through temporal IoU fusion~\cite{HUNGARIAN}. Fig.~\ref{fig:codec_results} compares eight configurations, comprised of model combinations presented in~\Cref{tab:setup}, across three deployment modes: cloud-only monolithic baselines that process every frame on a RTX 4090 GPU server, edge-only baselines that rely entirely on a lightweight model running on an Nvidia Jetson device, and Compound AI configurations that combine both. To isolate each pattern's contribution, we compare CODEC configurations against both baselines. The efficiency gain over cloud-only measures the Router's contribution, and the accuracy gain over edge-only measures the Aggregator's contribution.


\textbf{The Router reduces latency and cost via selective offloading.} Large monolithic baselines achieve the highest accuracy (Swin: 28.8\% mAP@50) but incur high latency (161~ms) and cost (\$0.018/1K frames). By offloading only selected frames, the Router enables CODEC configurations to reduce latency by up to 60\% and cost by 71\%. All compound configurations operate within 52--65~ms, confirming the efficiency of selective offloading.

\textbf{The Aggregator recovers accuracy lost by edge-only deployment.} Edge-only inference is fast (45--58~ms) and cheap (\$0.001--0.002/1K frames) but limited in accuracy. YOLOv8s alone reaches 21.0\% mAP@50, which is 7.8 points below the Swin cloud-only baseline. The Aggregator closes most of this gap by fusing asynchronous cloud predictions with continuous edge detections through a matching algorithm. Best CODEC configuration, YOLOv8s~+~Swin, achieves 25.5\% mAP@50, recovering 4.5 of those 7.8 points (58\%) at only 7~ms additional latency over edge-only baseline. Critically, none of the monolithic baselines reach this operating point. CODEC configurations occupy a position in the accuracy-latency-cost space that monolithic deployments cannot provide.

\textbf{Model selection governs the magnitude of these gains.} While the patterns offer potential for performance benefits, the choice of which model fills each role determines how much benefit is realized. Among Compound AI configurations, accuracy varies by 7.7 percentage points (17.8\% to 25.5\%). The selection of edge model has a substantial effect. YOLOv8s-based configurations outperform MobileNetV2-based ones by 4--6 percentage points regardless of cloud model. Since the edge model processes every frame while the cloud model processes only offloaded ones, its quality dominates overall system accuracy.

\begin{takeaway}
CODEC demonstrates that the Router and Aggregator patterns together provide an operating point that neither monolithic baseline can reach, combining accuracy close to the cloud model with latency close to the edge model. However, model selection determines the magnitude of these gains and plays a vital part in determining Compound AI performance. 
\end{takeaway}

\subsection{QARouter: Retrieval-Augmented Question Answering}\label{subsec:qarouter}

QARouter composes the Retriever and Router patterns for question answering. Every incoming query first passes through retrieval, where relevant passages are fetched from a FAISS index. A trained DistilBERT classifier~\cite{sanh2020distilbertdistilledversionbert} then evaluates query difficulty given the retrieved context and dispatches to either a local LLM or a proprietary model through an API. The router is trained following prior work~\cite{ding2024hybridllm}, using repeated small-model sampling on SQuADv2.0 training split, to estimate per-query difficulty and derive routing labels. 



\textbf{Routing reduces cost at the expense of accuracy.} Routing alone trades accuracy for cost reduction. The best routed configuration (GPT+Qwen) achieves 87.6 F1, 6.1 percentage points below the GPT-4o-mini monolithic baseline at 27\% lower cost. This gap establishes the baseline contribution of routing in isolation and motivates the addition of retrieval.

\begin{figure*}[t]
\centering
\includegraphics[width=0.9\linewidth]{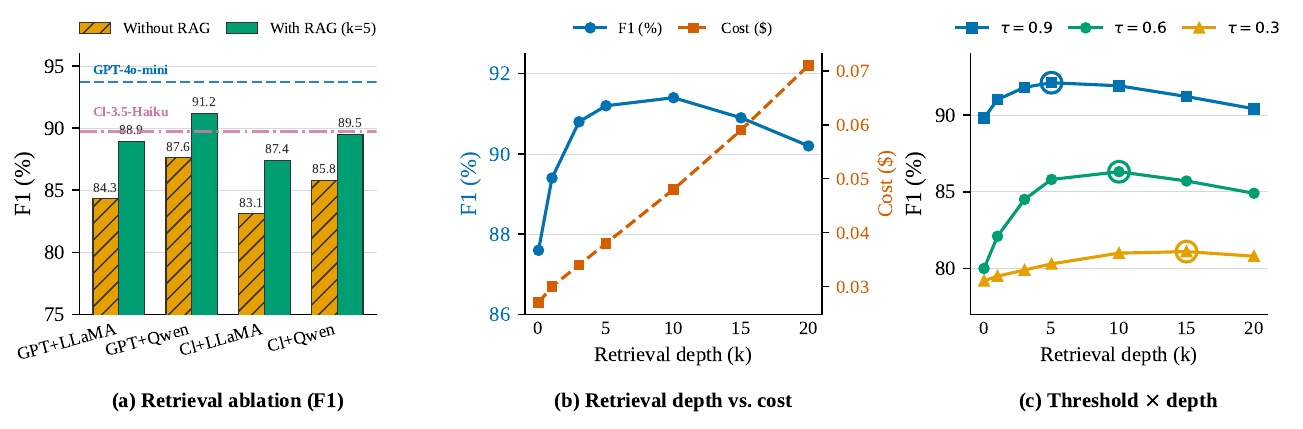}
\caption{QARouter performance on SQuAD dataset. (a)~Retrieval effect on F1. (b)~F1 \& cost as a function of retrieval depth. (c)~Parameter coupling effect on F1.}
\label{fig:qa}
\end{figure*}

\textbf{Retrieval improves accuracy across all configurations.} Fig.~\ref{fig:qa}(a) compares routed configurations with and without retrieval ($k{=}5$ passages). Adding retrieval 
improves F1 by 3.6--4.6 points across all model pairs, regardless of which proprietary model handles complex queries. GPT+Qwen with retrieval reaches 91.2 F1, within 2.5 points of the GPT-4o-mini monolithic baseline at 27\% lower cost. The consistency of the retrieval gain across model pairs confirms that the Retriever's contribution does not depend on the downstream models.

\textbf{Retrieval depth creates a non-linear accuracy-cost trade-off.} Fig.~\ref{fig:qa}(b) shows F1 and cost as a function of $k$ for GPT+Qwen configuration. Accuracy rises from $k{=}0$ to $k{=}3$ (+3.2 F1), plateaus between $k{=}5$ and $k{=}10$, then degrades beyond $k{=}10$ as irrelevant passages introduce noise into the generation context. Cost, by contrast, rises linearly with $k$ because each additional passage increases the prompt token count. At $k{=}5$, cost is 42\% higher than the no-retrieval baseline; at $k{=}20$, it is 163\% higher with lower 
accuracy than $k{=}5$. The optimal $k$ lies in the plateau region where accuracy has saturated but cost has not yet escalated. This represents a balance that cannot be identified without empirical evaluation across the full range.

\textbf{Retrieval depth and routing threshold are interacting parameters.}
Fig.~\ref{fig:qa}(c) shows that the optimal $k$, which maximizes accuracy, shifts with the routing threshold $\tau$, which determines the confidence required to dispatch a query to the local model. At $\tau = 0.9$, only queries classified as easy with high confidence by the router are processed locally. Consequently, the proprietary model handles the majority of traffic including many moderately easy queries. These queries are well-served by moderate context, and $k = 5$ is sufficient. At $\tau = 0.3$, most queries are dispatched locally and only the hardest queries reach the proprietary model. Hard queries benefit from richer context, shifting the optimum to $k = 15$. The magnitude of retrieval's contribution varies accordingly. At $\tau = 0.6$, retrieval improves F1 by 6.3 points, compared to 2.3 points at $\tau = 0.9$ and 1.9 points at $\tau = 0.3$. Tuning $k$ at a fixed $\tau$, or vice versa, yields suboptimal configurations, requiring joint optimization of both parameters. 


\begin{takeaway}
QARouter reaches monolithic baseline at 27\% lower cost, demonstrating that Compound AI can approach monolithic accuracy while improving efficiency. Retrieval depth produces a non-monotonic accuracy-cost relationship whose optimum cannot be determined by intuition alone. When the multiple patterns compose, the optimal retrieval depth shifts with the routing threshold, making independent parameter tuning insufficient and joint optimization necessary.

\end{takeaway}

\subsection{InferScale: Test-Time Compute Scaling}\label{subsec:inferscale}

InferScale instantiates the Sampler pattern for mathematical reasoning on GSM8K. The system generates $N$ candidate answers from LLaMA~3.2:3B using repeated sampling at 0.7 temperature and selects the final response through majority voting. The sample count $N$ is the primary configuration parameter.

\begin{figure}[h]
\centering
\includegraphics[width=0.9\columnwidth]{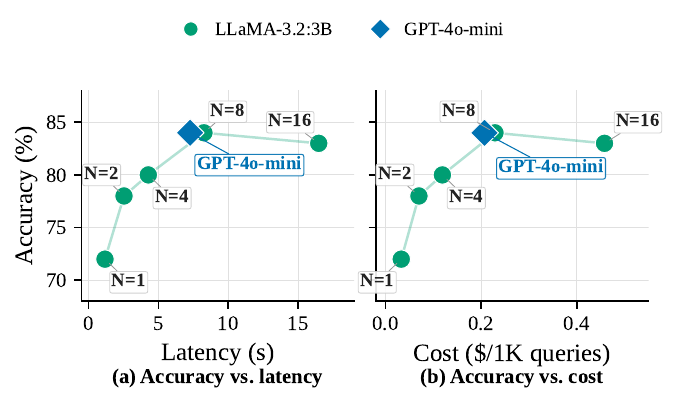}
\caption{InferScale performance on GSM8K.}
\label{fig:inferscale_results}
\end{figure}

\textbf{Repeated sampling closes the accuracy gap to a larger model at lower cost.} \Cref{fig:inferscale_results} plots each configuration in the accuracy-latency plane~(a) and the accuracy-cost plane~(b). Increasing $N$ moves the operating point along the planes. At $N{=}4$, the system reaches 80\% accuracy, within 4 percentage points of the GPT-4o-mini baseline, while reducing cost by 43\% and latency by 41\%. This configuration demonstrates that the Sampler pattern can approach monolithic accuracy at substantially lower operational cost by allocating inference-time compute within a smaller local model rather than delegating to a larger one. At $N{=}8$, accuracy matches the GPT-4o-mini baseline at 84\%, confirming that the pattern can reach the monolithic ceiling when accuracy is the binding constraint.

\textbf{Accuracy degrades beyond the optimal sample count.} At $N{=}16$, accuracy drops to 83\% while latency doubles to 16.5~s and cost rises to \$0.458 per 1K~queries. Frequent but incorrect answers begin to dominate the majority vote. This configuration is strictly dominated by $N{=}8$, which achieves higher accuracy at half the latency and cost. Additional compute, therefore, does not unconditionally improve quality, and the optimal $N$ must be determined empirically.

\begin{takeaway}
The Sampler pattern offers a potential for a smaller model to approach monolithic accuracy through repeated sampling and majority voting. The accuracy-efficiency relationship is non-linear. Beyond optimal sample size, additional samples degrade both accuracy and efficiency.    
\end{takeaway}

\subsection{Discussion}\label{subsec:discussion}

\begin{table}[h]  
\centering  
\caption{Compound vs.\ monolithic baselines. Values show relative change; negative latency and cost indicate savings.}  
\label{tab:summary}  
\footnotesize  
\renewcommand{\arraystretch}{1.2}  
\begin{tabular}{@{}l l r r r@{}}
\toprule
\textbf{Case Study} & \textbf{Pattern} & \textbf{Accuracy} & \textbf{Latency} & \textbf{Cost} \\
\midrule
CODEC      & Router+Agg.      & $-$3.3\,pp  & $-$60\%  & $-$71\%    \\
QARouter   & Router+Retriever & $-$2.5\,pp  & $-$21\%  & $-$27\%    \\
InferScale & Sampler          & $-$4\,pp    & -41\%    & $-$43\%  \\
\bottomrule

\end{tabular}
\end{table}  

Table~\ref{tab:summary} summarizes Compound AI performance relative to monolithic baselines across the three case studies. These results establish four findings about Compound AI design trade-offs.

\textbf{Compound AI approaches monolithic accuracy while reducing latency and cost, opening operating points monolithic deployments cannot provide.} Section~\ref{sec:motivation} showed that each model in Fig.~\ref{fig:monolithic} occupies a single fixed point in the accuracy-latency-cost space, with no mechanism to trade one objective against another at runtime. The case studies break this property. CODEC achieves 25.5\% mAP@50 at 65~ms, which no cloud or edge model alone reaches, since cloud models require 157--161~ms and edge models peak at 21.0\%. QARouter and InferScale likewise come within 2.5 and 4 percentage points of their monolithic baselines at lower cost, each occupying a position that Fig.~\ref{fig:monolithic} shows is inaccessible to monolithic deployment.

\textbf{Model selection determines the magnitude of Compound AI gains.} While the design patterns create structural potential for trade-off management, the choice of which model fills each role determines how much of that potential is realized. In CODEC, edge model selection has the highest impact on the overall accuracy because the edge model processes every frame, while the cloud model processes only the offloaded fraction. YOLOv8s-based configurations outperform MobileNetV2-based ones by 4--6 percentage points regardless of utilized cloud model. In QARouter, the choice of local and proprietary model pair shifts F1 score by up to 4.5 percentage points across configurations. 

\textbf{Parameter configuration shapes the trade-off space non-linearly.} Within a fixed model assignment, parameter values determine where the system operates in the accuracy-latency-cost space. However, the relationship between parameters and system objectives is non-monotonic. In InferScale, increasing the sample count from $N=1$ to $N=8$ improves accuracy from 72\% to 84\%, but further increasing to $N=16$ degrades accuracy to 83\% while doubling latency and cost. Additional inference-time compute does not unconditionally improve quality. In QARouter, retrieval depth beyond $k=10$ degrades accuracy as irrelevant passages introduce noise into the generation context, while cost continues to rise linearly with $k$. In both cases, the optimal parameter value lies in a region that cannot be identified by intuition alone and must be determined empirically.

\textbf{Model selection and parameter configuration interact across composed patterns.} The case studies show that these configuration choices are interdependent. In QARouter, the routing threshold determines which model handles which queries. As a consequence, it affects how much each model's quality contributes to overall F1, and what retrieval depth is optimal, as more capable models typically require less context size. This means that Compound AI trade-offs cannot be managed by optimizing each configuration choice in isolation. As systems compose more patterns, these interactions multiply, making manual configuration increasingly unreliable.

\section{Research Roadmap}\label{sec:challenges}

The case studies demonstrate that Compound AI configurations can achieve operating points unattainable by monolithic deployments. However, every configuration evaluated in \Cref{sec:trade} was discovered through systematic but tedious, time-consuming and costly manual experimentation. We argue that the long-term goal for Compound AI systems is enabling automatic identification of optimal configurations while maintaining SLO compliance as production conditions change. We identify five key future research challenges (RC) that should be addressed in pursuit of this goal.

\textbf{RC-1: Automatic Design Space Exploration}. A Compound AI workflow exposes many design choices, including model selection for each inference task, the algorithms and tools assigned to each workflow stage, and parameters that govern runtime behavior. Even workflows modest in size can produce thousands of candidate configurations. Gradient-based optimization cannot navigate this space because the workflows are non-differentiable. Components are heterogeneous and many choices are discrete or categorical, such as routing decisions or model assignments. Black-box methods offer an alternative, but the space is not decomposable. As Section IV demonstrated, parameters interact across composed patterns, meaning that each candidate configuration must be evaluated jointly through a full end-to-end workflow execution. This makes evaluation expensive and rules out independent per-component tuning. How can such configuration space be searched efficiently when both gradient-based and decomposed search strategies are inapplicable, is an open research challenge.

\textbf{RC-2: Dynamic Runtime Adaptation.}
The configurations evaluated in Section~\ref{sec:trade} were static, optimized offline for a fixed dataset and infrastructure. However, production conditions change. For instance, query distributions can shift, model APIs update pricing or deprecate versions, and hardware load fluctuates with demand. Under any of these shifts, a previously optimal configuration can drift outside the SLO-compliant trade-off space. Restoring compliance may require adjustments at different granularities, such as tuning parameter values, runtime model selection and switching, or restructuring the workflow topology. We envision novel adaptation mechanisms that can dynamically adjust system configuration to restore and maintain SLO compliance as conditions change.

\textbf{RC-3: Novel Programming Abstractions.} Runtime adaptation requires reassigning models to inference tasks without disrupting workflow execution, but models performing the same task often differ in interfaces, output formats, and invocation protocols. Current Compound AI frameworks bind each inference task to a specific model, so reassigning a model at runtime risks breaking downstream workflow steps that depend on the previous model's output format. Novel programming abstractions are needed that decouple inference tasks from the models assigned to them, providing stable interfaces so that model reassignment does not disrupt workflow execution.


\textbf{RC-4: Serverless Compound AI Systems.} The adaptation mechanisms above require that runtime model selection is cheap, but conventional deployment ties models to GPU memory and serving infrastructure, making switching expensive. A FaaS execution model with scale-to-zero capabilities eliminates the need to keep candidate models pre-loaded in memory, making model switching a lightweight operation rather than an infrastructure-level disruption. However, Compound AI workflows are multi-stage pipelines, and executing them as compositions of stateless functions raises open problems in intermediate state management, inter-stage communication, and scheduling that respects end-to-end workflow latency rather than optimizing individual invocations.

\textbf{RC-5: Quality Attribution in Compound AI Workflows.} When overall task accuracy degrades in a Compound AI workflow, the root cause may lie in any upstream model or component. Consequently, the error propagates through the workflow in ways that infrastructure-level metrics, such as cost or request latency, cannot reveal. The adaptation mechanisms from RC-2 require knowledge about which component to adjust, yet pinpointing where in the workflow the quality degradation originates remains an open problem. Estimating per-component accuracy contribution in a non-differentiable system is challenging given that components interact non-linearly, errors compound across stages, and there is no gradient signal to trace quality loss back to its source. Solving the quality attribution problem is one of the main enablers for the adaptation mechanisms envisioned in RC-2.



These five challenges define the research roadmap from manually configured Compound AI prototypes towards systems that can automatically discover and maintain configurations that satisfy SLO constraints. RC-1 concerns finding good configurations offline. RC-2 through RC-5 concern keeping the system within its SLO-compliant operating region once deployed. We consider that advancing Compound AI to production scale requires progress across all five directions.

\section{Related Work}\label{sec:related}

Our work intersects three lines of research: AI paradigms that compose multiple components, prior work on Compound AI systems, and distributed systems whose architectural principles we build upon but whose trade-off space we redefine around task performance.

\subsection{Related AI Paradigms}

\subsubsection{Agentic AI}
Agentic AI refers to systems in which LLM-driven agents autonomously decide which actions to take, which tools to invoke, and in what sequence~\cite{acharya2025agentic, schneider2025generative}. Foundational work such as ReAct~\cite{yao2023react} demonstrated that interleaving reasoning traces with tool-calling actions enables LLMs to solve multi-step tasks without a predefined execution plan. Multi-agent extensions coordinate several such agents through communication protocols or shared state~\cite{guo2024multiagent}, and recent position papers argue that understanding these systems requires a systems-theoretic perspective that accounts for emergent behavior arising from agent interactions~\cite{miehling2025systems}.

Key distinction between agentic and Compound AI lies in workflow control. In agentic systems, the workflow is \textit{implicit}: it emerges at runtime from the agent's reasoning and planning~\cite{yao2023react}. In Compound AI, the workflow is \textit{explicit}: practitioner specifies which component process which inputs, in what order, and under what conditions. Explicit workflow structure exposes a bounded design space in which the effect of each decision on system objectives can be optimized.

\subsubsection{Modular AI}
Modular AI architectures compose specialized sub-networks into task-specific computation graphs. Mixture-of-experts models~\cite{shazeer2017moe} route inputs to specialized sub-networks within a single model, while modular deep learning~\cite{pfeiffer2024modular} assembles independently trained modules. These systems are trained end-to-end via backpropagation because all components are differentiable.

Compound AI systems break this assumption. Due to heterogeneity of building blocks and design patterns in Compound AI, these systems are non-differentiable~\cite{LLMSELECTOR}, which prevents gradient-based joint optimization. Recent work addresses this through surrogate methods, such as approximate gradients~\cite{TEXTGRAD, COMPASS}, system-level preference optimization~\cite{wang2025sysdpo}, and local reward functions aligned with global performance~\cite{OPTIMAS}. Our work is complementary. We provide the architectural abstractions that define the design space over which such methods operate.




\subsection{Classical Distributed Systems}

Compound AI systems are distributed systems. They coordinate heterogeneous components across tasks, deployments, and network boundaries. Classical distributed systems research provides foundational principles for service composition, fault tolerance, and scheduling~\cite{dean2008mapreduce, newman2021microservices} that Compound AI systems inherit. However, classical distributed systems optimize for infrastructure-level objectives, such as throughput, availability, and resource utilization. Compound AI systems introduce \textit{task performance} (e.g., accuracy) as a first-class system objective that interacts with infrastructure objectives in that classical systems do not exhibit. A routing decision in a load balancer optimizes latency or throughput, while a routing decision in a Compound AI system trades accuracy against cost, as different models produce outputs of different quality for the same input. Our design methodology operates at this boundary, presenting Compound AI design as workflow topology and configuration selection, making the trade-offs between task performance and operational efficiency manageable.

\section{Conclusion}\label{sec:conclusion}

We presented a design methodology for Compound AI systems. Methodology is organized through workflow topology design, captured through eight design patterns that each address a specific limitation of monolithic deployment and configuration selection, which assigns models to inference tasks within the workflow and configures parameters that govern runtime behavior. These dimensions define a structured design space for building AI systems to meet service-level objectives that no monolithic model can satisfy independently.

Three case studies across vision and language domains demonstrate that Compound AI systems can approach monolithic baselines in task performance while improving efficiency. CODEC reaches within 3.3 percentage points of the monolithic baseline for object detection while reducing latency by 60\% and cost by 71\%. QARouter matches 97\% of monolithic accuracy at 27\% lower cost. InferScale enables a 3B-parameter model to match GPT-4o-mini on mathematical reasoning through repeated sampling. Across the studies, model selection dominates system-level performance and configuration parameters interact non-linearly, requiring joint optimization.

These findings are based on manual experimentation over research-scale testbeds with established benchmarks. Future work will focus on addressing research challenges outlined in \Cref{sec:challenges} to advance from manual prototyping towards Compound AI systems that automatically discover and maintain SLO-compliant configurations in production environments.

\section*{Acknowledgment}

This work was partly funded by the European Union under the Horizon Europe programme through the SNS JU (Grant Agreement No. 101192912, NexaSphere). Views expressed are those of the authors and do not necessarily reflect those of the EU or the SNS JU.

\bibliographystyle{IEEEtran} 
\bibliography{references} 

\end{document}